\documentclass[pra,amssymb,amsmath,showpacs]{revtex4}

\usepackage{graphicx}
\usepackage{amssymb}
\usepackage{amsmath}
\usepackage{color}

\newcommand{\be}{\begin{equation}}
\newcommand{\ee}{\end{equation}}
\newcommand{\bea}{\begin{eqnarray}}
\newcommand{\eea}{\end{eqnarray}}

\begin{document}       

\title{How Does the Brain Organize Information?}

\author{H.~Kr\"{o}ger$^{a}$$\footnote{Corresponding author, 
Email: hkroger@phy.ulaval.ca}$} 

\affiliation{
$^{a}$ {\small\sl D\'{e}partement de Physique, Universit\'{e} Laval, Qu\'{e}bec, Qu\'{e}bec G1K 7P4, Canada} \\ 
\\ \ \\
Original version, \today 
}

\begin{abstract}
Cognitive processes in the brain, like learning, formation of memory, 
recovery of of memorized images, classification of objects (like is this object a table or not?) have two features: First, there is no supervisor in the brain who controls these processes (like is the memorized image of a person true?). Second there is a hugh number of neurons ($10^{6}$ to $10^{10}$) involved in those cognitive tasks.
For this reason, the search of understanding cognitive processes 
uses models built from a large number ($10^{2}$ to $10^{6}$) neurons, but very much simplified neurons. The so-called neural networks have been quite successful describing certain aspects of brain functions, like the mechanism of associative memory or recently the prediction of epileptic seizures. At hand of the Kohonen network we discuss
the treatment of information in the brain, in particular how the brain organizes such information without supervisor. Recently, networks of small-world and scale-free architecture came into focus. There is evidence indicating that the brain (Cat cortex, Macaque cortex, human brain) uses such connectivity architecture. 
Tasks like treatment of information, learning and classification take advantage of such scale-free and small-world connectivity and thus play a potentially important role in the self-organization of the brain.
\end{abstract}

\pacs{} 

\maketitle

\section{Introduction: Overview of organizational principles}
\label{sec:Intro}
An important principle of neural organization of the brain are the so called feature maps. This can be (i) sets of feature-sensitive cells, (ii) ordered projections between neuronal layers or (iii) ordered maps of abstract features. An example of the first kind are cells responding to human faces. Examples of the second type is the retinotopic map of the visual field or the somatotopic 'homunculus' in the sensorimotor cortex. Examples of the third type is the color map in the visual area V4, the directional hearing map in the owl midbrain~\cite{Knudsen78} and the target-range map in the bat auditory cortex~\cite{Suga79}. More than one response property can be mapped simultaneously within an area, as has been shown in visual, somatosensory and auditory cortex~\cite{Hubel63,Sur81,Linden03,Friedman04}.
As no receptive surface exists for such abstract features, the spatial order of representations must be produced by some self-organizing process in brain, which occurs mainly postnatally. 
It has been proposed that those maps need to smoothly map several response properties onto a two-dimensional cortical surface (dimensional reduction), which has led 
Kohonen~\cite{Kohonen82} and later Durbin et al.~\cite{Durbin90}, 
Obermayer et al.~\cite{Obermayer90} and Swindale~\cite{Swindale91} to propose models of formation of such feature maps. 
The biological relevance of Kohonen model has been established 
in a number of cases. Obermayer et al.~\cite{Obermayer90,Obermayer92}, 
Goodhill~\cite{Goodhill93}, Wolf et al.~\cite{Wolf94}, Swindale and Bauer~\cite{Swindale98} have shown that the Self-Organized Map (SOM,Kohonen model) is able to produce mappings that bear a detailed resemblance to real visual cortex maps. Swindale et al.~\cite{Swindale00} analyzed visual cortex maps in cat and found coverage optimization important for cortical map development, lending support to dimension reduction models of Kohonen~\cite{Kohonen82} and Durbin et al.~\cite{Durbin90}. 
Swindale~\cite{Swindale04} used the Kohonen model to study how many different feature spaces may be represented in cortical maps. 
SOM have been used to study attentional impairment and familarity preference in autism 
\cite{Gustafsson04}. By investigation of the 
primary visual cortex (V1) of ferret Sur et al.~\cite{Sur05} found a distorsion in the mapping of the visual scene onto the cortex, in agreement with the predictions of the Kohonen model. Aflalo et al.~\cite{Aflalo06} investigated origins of complex organization of motor cortex and found that the map generated from the Kohonen model contained many features of actual motor cortex in monkey.

It is a biological property that the organization of cortex in mammals occurs during a short time window after birth during which neural connectivity is established. Most remarkably, this is accompanied by genetically controlled neuron cell death and pruning of synaptic 
connections~\cite{Shepherd94, Kandel95}. Post-natal pruning of synaptic connections is a biological principle of organization of brain.

Another organizational principle potentially lies in the neural connectivity architecture and its resulting brain function. The concept of Small World Networks (SWN)~\cite{Watts98} and the variant of Scale Free Networks (SFN)~\cite{Albert99} has become very popular recently, after the discovery that such architecture of network wiring is realized in the organization of human society (Milgram's experiment), in languages, in the WWW, in the internet, in the metabolic network of E. coli, in the nervous system of the nematode worm C. elegans, in cat cortex, and in macaque cortex.  In brains of humans, functional magnetic resonance imaging (FMRI) has shown evidence for functional networks of correlated brain activity~\cite{Eguiluz05}, which display SWN and SFN architecture.
On the other hand, $1/f^{\alpha}$ frequency scaling has been observed in electric local field potentials~\cite{Bedard06b} hinting to long ranged temporal correlations.
The possibility that SWN and SWF architecture plays a role in the organization of brain leads
the important question: Does SWN and SFN architecture bring about a functional advantage in the working of the brain? 
In fact, a Hodgkin-Huxley network with SWN wiring has been found to give a fast and synchronized response to stimuli in the brain~\cite{Lago00}. Associative memory models have shown that SWN architecture yields the same memory retrieval performance as randomly connected networks, however, using only a fraction of total connection length~\cite{Bohland01}. Modeling supervised learning in a layered feed-forward network with SWN architecture was found to reduce learning time and error~\cite{Simard05}. 
These findings support the hypothesis that SWN/SFN architecture may represent another organizational principle of brain.

Electrophysiological experiments have reveiled power laws in the frequency behavior of electrical signals in brain.
The electroencephalogram (EEG) and magnetoencephalogram display
frequency scaling close to $1/f$~\cite{Pritchard92,Novikov97}. EEG
analysis~\cite{Link01} and avalanche analysis of local field
potentials (LFPs) recorded {\it in vitro}~\cite{Beggs03} showed power-law
distributions. Also power-law behavior has been found in interspike interval (ISI) distributions computed from retinal, visual thalamus and primary visual cortex
neurons~\cite{Teich97}. In the quest for an underlying common principle, 
the sand-pile model of self-organized criticality~\cite{Bak87} has been proposed to be 
at work in the brain.
Self-organized critical states are found for many complex systems in
nature, from earthquakes to avalanches~\cite{Bak96,Jensen98}. Such
systems are characterized by scale invariance, which is usually
identified as a power-law distribution of variables such as event
duration or the waiting time between events. $1/f$ noise is usually
considered as a footprint of such systems~\cite{Jensen98}. $1/f$
frequency scaling indicates long-lasting correlations in the system, 
similar to the behavior of condensed matter physical systems near critical
points (phase transitions).

In a study of self-organization of young brain based on the Kohonen model or Self Organized Map (SOM)~\cite{Kohonen82}, Pallaver~\cite{Pallaver06} has asked if several of the above organizational principles may be at work at the same time, possibly being interconnected.
He found that (i) introducing adaptive learning rates improves the final error of the feature map, (ii) SWN topology is present during most of the organizational phase making the response to stimuli more efficient, and (iii) pruning of synaptic connections in the early evolution of the network is beneficial for organization, while reconnections give an adverse effect. While the latter study can be considered as a first step to understand the working and interplay of organizational principles in the cortex, there are a lot of open questions and possible extensions to be investigated at hand of the Kohonen model: (1) In the Kohonen model the number of connections is gradually reduced. Does the observed SWN architecture persist until connectivity is diluted to a degree observed biologically in brain (very sparce connectivity)? Does SFN architecture play a role? (2) Pruning of connections is related to plasticity of brain. How would any change of the rule of pruning modify the resulting organization? (3) Presently, in neuroscience much work is focusing on the role of glial cells in brain function. An interesting question is: Is the biological manifestation of the Kohonen map given in terms of neural connections, in terms of glial connections or rather neural-glial interconnections via synapses? How would the functioning of the Kohon map change by invoking beyond neural connections also glial cells, its connections and interaction with neural synapses?

\section{Sand pile model of self-organized criticality at work in the brain}
\label{sec:SandPile}
The sand pile model of self-organized criticality is an attempt to understand emergent complex behavior in physical and biological systems. The term emergent complexity denotes phenomena which can not be explained and predicted from properties of its elementary components. An example is the formation of snow flakes. Each snow flake is individually different. The form of a particular snow flake can not be predicted from the laws of the hydrogen and oxygen atoms. Such emergent complex behavior is in contradiction to the reductionist point of view of nature, which claims that the laws of nature can be understood, when the laws of all of its elementary components are understood. There is a large number of phenomena of emergent complexity in nature, a lucid presentation with many examples is given by 
Laughlin~\cite{Laughlin05}. A candidate of emergent complex behavior is the so-called $1/f$ noise. This means that a time-dependent signal $s(t)$ displays fluctuations which in terms of frequency $f$ behave like $1/f$. Such $1/f$ noise has been observed in nature in a large variety of phenomena. Examples are: fluctuations in electrical resistance of a conductor, sunspot activity, the flow of the river Nile, or pressure variations in the air caused by playing musical instruments. In order to explain the puzzle of $1/f$ noise, Bak, Tang and Wiesenfeld~\cite{Bak87,Bak96,Jensen98} have proposed the so-called sand pile model of Self-Organized Criticality (SOC). The sand pile model is a mathematical model (cellular automaton) describing in terms of simplified rules the motion of grains of sand and also avalanches of many grains on the surface of a pile of sand. The interesting feature is the occurence of avalanches and its statistical behavior in time and in size. The model predicts temporal $1/f$ 
(power law) behavior for the occurence of avalanches, which shows up in the power spectrum of the signal. Also the spatial behavior, i.e. size of avalanches, is predicted to have fractal behavior (power law behavior). Bak et al.~\cite{Bak87} showed that this means an absence of any scale in time and in space (in the function 
$g(x)=\exp[-x/a]$ with $x$ of dimension length, the parameter $a$ of dimension length sets a scale, however the function $h(x) = x^{\alpha}$ has no scale). The system has a meta-stable (critical) state, meaning that avalanches may be set in motion by a tiny perturbation of the system (in a real sand pile this may correspond to dropping a single grain of sand on top of the pile). Later, experiments with real sand piles have shown that the predictions of the model are not in good agreement with nature. However, Frette et al.~\cite{Frette96} have shown that piles of elongated rice grains behave according to the prediction of the SOC model. Self-organized critical states are found for many complex systems in nature, for example in avalanches, earthquakes, spreading of forrest fires, avalanches of electrons in superconductors of type II, water droplets formed on a window pane and biological evolution~\cite{Jensen98}. Such systems are all characterized by scale invariance, which is usually
identified by a power-law distribution of variables such as event
duration or the waiting time between events.  
\begin{figure}[ht]
\begin{center}
\includegraphics[scale=0.7,angle=0]{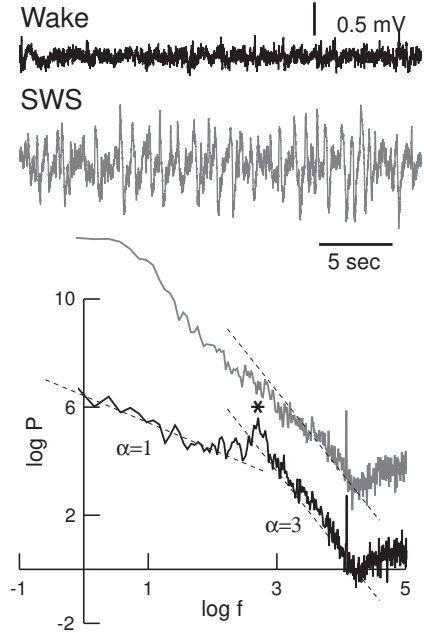}
\end{center}
\caption{Frequency scaling of local field potentials from cat
parietal cortex.  Top traces: LFPs recorded in cat parietal
cortex during wake and slow-wave sleep (SWS) states.  Bottom:
Power spectral density of LFPs, calculated from 55~sec sampled at
300~Hz (150~Hz 4th-order low-pass filter), and represented in
log-log scale (dashed lines represent $1/f^\alpha$ scaling). During
waking (black), the frequency band below 20~Hz scales approximately
as $1/f$ (*: peak at 20~Hz beta frequency), whereas the frequency
band between 20 and 65~Hz scales approximately as $1/f^3$.  During
slow-wave sleep (gray; displaced upwards), the power in the slow
frequency band is increased, and the $1/f$ scaling is no longer
visible, but the $1/f^3$ scaling at high frequencies remains
unaffected. PSDs were calculated over successive epochs of 32~sec,
which were averaged over a total period of 200~sec for Wake and
500~sec for SWS.}
\label{lfps}
\end{figure}
There are experiments giving evidence of the existence of such critical
states also in brain activity. Global variables, such as the
electroencephalogram (EEG) and magnetoencephalogram (MEG), display
frequency scaling close to $1/f$~\cite{Pritchard92,Novikov97}. EEG analysis~\cite{Link01} and avalanche analysis of local field potentials (LFPs)~\cite{Beggs03} provided evidence for self-organized critical states with power-law
distributions. There is also evidence for critical states from the
power-law scaling of interspike interval (ISI) distributions computed
from retinal, visual thalamus and primary visual cortex
neurons~\cite{Teich97}. In a neural network model of brain plasticty 
de Arcangelis et al.~\cite{Arcangelis06} found 
that critical states may be associated with frequency
scaling consistent with experiments. 
Moreover, $1/f$ spectra in a dissipative SOC model have been found~\cite{LosRios99}.
\begin{figure}[ht]
\begin{center}
\includegraphics[scale=0.7,angle=0]{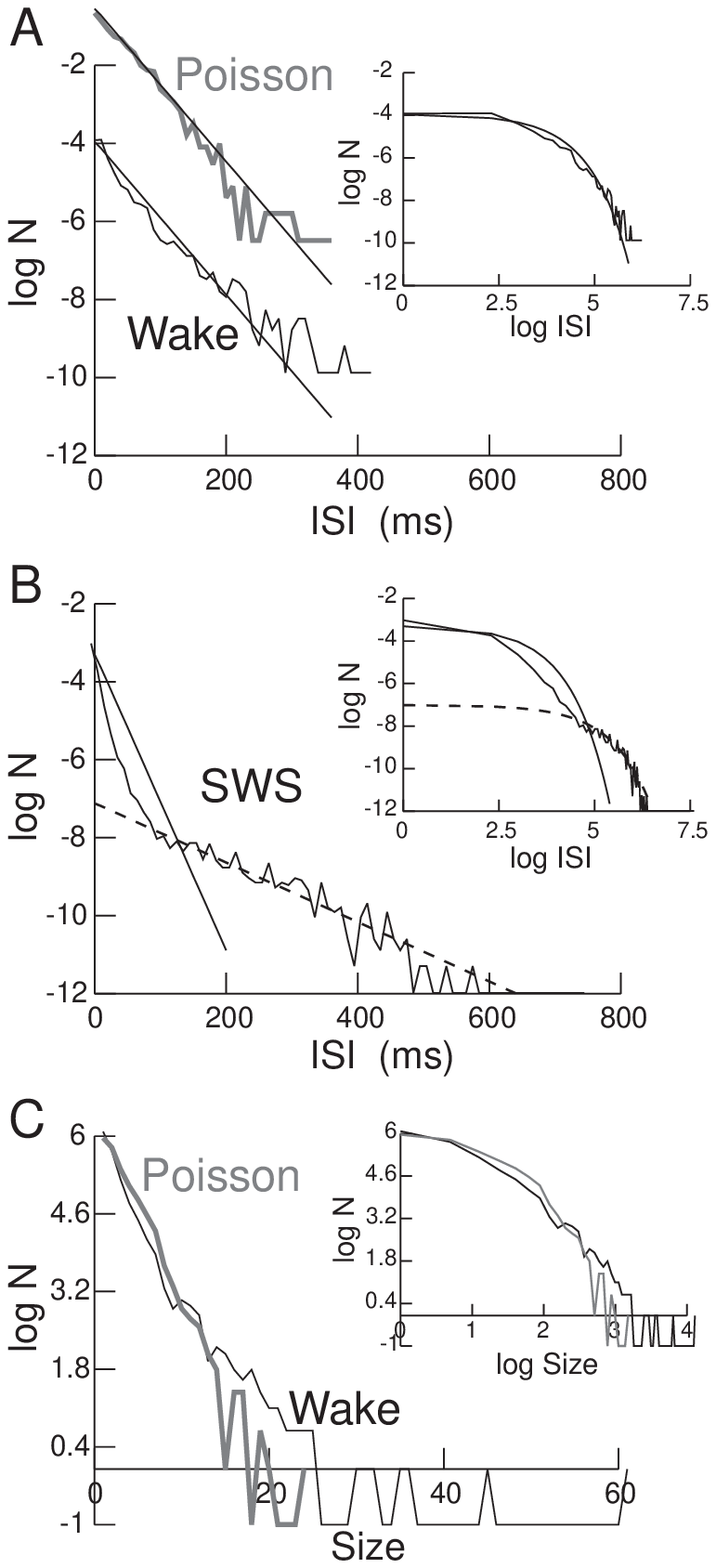}
\end{center}
\caption{Absence of power law distributions in neuronal activity. 
The logarithm of the distribution of interspike intervals (ISI)
during waking (Wake, A, 1951 spikes) and slow-wave sleep (SWS, B,
15997 spikes) is plotted as a function of ISI length, or log ISI
length (insets).  A poisson process of the same rate and statistics
is displayed in A (Poisson; gray curve displaced upwards for
clarity).  The exponential ISI distribution predicted by Poisson
processes of equivalent rates is shown as straight lines (smooth
curve in inset).  The dotted line in B indicates a Poisson process
with lower rate which fits the tail of the ISI distribution in SWS. 
C. Avalanche analysis realized by taking into account the
statistics from all simultaneously-recorded cells in Wake.  The
distribution of avalanche sizes scales exponentially (black curves),
similar to the same analysis performed on a Poisson process with same
statistics (gray curves).} 
\label{scale}
\end{figure}
However, in order to firmly establish the the SOC model as underlying mechanism of 
$1/f$ scaling in the brain, it would require to establish the presence of long-living meta-stable states~\cite{Paczuski96}. 
This has not been experimentally achieved yet. Thus the presence of the SOC mechanism has  not been firmly determined yet. In an attempt to answer these questions, 
B\'edard et al.~\cite{Bedard06b} investigated if $1/f$ frequency scaling is present in global variables recorded {\it in vivo} close to the underlying neuronal current sources (LFPs). They analyzed cortical LFPs which were recorded within cerebral cortex using
bipolar extracellular high-impedance
microelectrodes~\cite{Destexhe99}. LFP recordings sample
localized populations of neurons. This is in contrast to the EEG, which samples much larger populations of neurons~\cite{Niedermeyer98} and is being recorded from the surface of the scalp using millimeter-scale
electrodes. LFP signals undergo much less filtering than EEG signals,
because the latter scatter through various media, like
cerebrospinal fluid, dura matter, cranium, muscle and skin. Thus,
finding $1/f$ frequency scaling of bipolar LFPs would be a much
stronger evidence that this scaling reflects neuronal activities, as
these signals are directly recorded from within the neuronal tissue. 
In order to distinguish state-dependent scaling properties,
data have compared recordings during wakefulness and slow-wave sleep in
the same experiments~\cite{Bedard06b}.   

The following observations were made. LFPs from cat parietal association cortex show typical landmarks of EEG signals~\cite{Steriade03}. 
During waking, LFP signals have low amplitudes and display very irregular (chaotic) temporal behavivor (Fig.~\ref{lfps} top). They are dominated by low frequencies around 20~Hz ($\beta$ frequencies).
This pattern of ``desynchronized'' activity is typically seen during aroused states in human EEG~\cite{Niedermeyer98}. During slow-wave sleep, LFPs show
high-amplitude low-frequency activity (Fig.~\ref{lfps} middle),
similar to the ``$\delta$ frequencies'' of human sleep
EEG~\cite{Niedermeyer98}. The power spectral density (PSD)
obtained from these LFPs (Fig.~\ref{lfps} bottom) 
shows a broad-band structure. 
During wakefulness, one observes in the PSD the existence of 
two different scaling windows. For low frequencies ($1~Hz < f < 20~Hz$)
the PSD scales approximately like $1/f$, whereas for higher
frequencies ($20~Hz < f < 65~Hz$) the PSD scales approximately like
$1/f^3$ (Fig.~\ref{lfps}, black PSD). During slow-wave sleep, the
same $1/f^3$ scaling is observed in the high-frequency band
(Fig.~\ref{lfps}, gray PSD). These results are consistent with the findings of $1/f$ frequency scaling in the EEG~\cite{Pritchard92}, but for cat association cortex LFPs this is observed only during waking and for specific frequency bands.

The finding of a window of $1/f$ frequency scaling in the PSD of cat cortex LFP's reflects self-similarity in the time-evolution of signals. Because the PSD is essentially the same as the auto-correlation function, this indicates some long range correlation. It may be a signal for the presence of SOC. However, as discussed above, SOC requires the presence of a meta-stable state, which is difficult to detect experimentally. However, in the presence of SOC one would expect self-similarity and a power law behavior not only in the distribution of frequencies but also in the distribution of the inter-spike-intervals (ISI) as well as in the distribution of the size of avalanches. 
This has been analyzed in Fig.~\ref{scale}. One observes that the distribution of ISIs follows a Poisson distribution and not a power law distribution (Fig.~\ref{scale} top and middle). The same holds for the distribution of the avalanche sizes (Fig.~\ref{scale} bottom). This evidence is inconsistent with the SOC model.

What is the explanation of the dynamical behavior of $1/f$ and $1/f^{3}$ scaling windows observed in LFPs of cat cortex? B\'edard et al.~\cite{Bedard04,Bedard06a} have studied models of propagation of electrical signals in brain tissue. In Ref.~\cite{Bedard06a} in a model considering neurons as electrical sources and glial cells standing for a responsive  medium, and taking into account resistive and capacitive properties, the model predicted attenuation of high frequencies, consistent with experiments. Moreover, the model was found to be electrically equivalent (in the case of a single neuron source and a single passive glial cell) to an RC-circuit. This suggests that a system of many neurons and many glial cells may be electrically equivalent to a system of coupled RC-circuits. It has been shown by Barnes and 
Jarvis~\cite{Barnes71} that a large (infinite) chain of coupled RC-cicuits generates $1/f$ noise. This may give a possible explanation of $1/f$ frequency scaling observed in LFPs of cat cortex, as being a property of the brain tissue. The scaling window showing $1/f^{3}$ scaling may be due to synaptic noise (showing $1/f^{2}$ scaling) convoluted with $1/f$ noise of the brain tissue.

In conclusion of this section it has been shown that the PSD of bipolar LFPs from cat
parietal cortex displays several scaling regions, behaving like $1/f$ or $1/f^3$
depending on the frequency band and behavioral state.  The analysis
of neuronal unit activity from the same experiments does not support the hypothesis
that $1/f$ scaling in LFPs is associated with critical
states. Neither ISI distributions nor avalanche size distributions
show power-law scaling, but are rather consistent with Poisson
processes. An explanation for $1/f$ frequency scaling, alternative to the SOC model, has been suggested, which is based on the filtering properties of extracellular media.  These results may appear to contradict previous evidence for critical states
in vitro~\cite{Beggs03} or in the early visual system in
vivo~\cite{Teich97}. However, the absence of critical states
reported here may instead reflect fundamental differences between
association cortex and other structures more directly related to
sensory inputs.

\section{Neural Connectivity in Brain: Role of Small World and Scale Free Architecture}
\label{sec:SWNTopology}
The concept of a Small World Network (SWN) has been invented by Watts and Strogatz~\cite{Watts98}. It explains Milgram's letter experiment which shows that on average six persons form a chain of personal connections which links some person on earth to any other person on earth (the original experiment was done by sending letter from some place in mid-west USA to Boston).
This letter experiment says something about the orginization of human society. 
Actually it turnes out that such SWN architecture is realized in various ways in nature, like in the WWW~\cite{Albert99,Huberman99}, in the
internet~\cite{Faloutsos99}, in the distribution of electrical power
(western US)~\cite{Watts98}, or in the metabolic network of the
bacterium {\it Escherichia coli}~\cite{Strogatz01,Wagner01}. 
Shortly afterwards a variant of SWN has been proposed by 
Barab\'{a}si et al.~\cite{Albert99,Barabasi99,Albert00,Jeong00}, the so called Scale Free Network (SFN). Recent reviews on SWN can be found in Ref.~\cite{Strogatz05} and on SCF in 
Ref.~\cite{Barabasi03}.

An SWN is characterized by the following properties. It is a network with high local clustering, that is locally close nodes are likely to be connected 
(mathematically described by a clustering coefficient $C$ which measures the
probability that given node $a$ is connected to nodes $b$ and $c$ then also 
nodes $b$ and $c$ are connected). Second, the SWN has a small average path length  
(mathematically is is expressed via a function $L$, and small $L$ means geometrically distant nodes are connected sometimes by very few intermediate nodes, i.e. more or less directly). As result, $C$ is large and $L$ is small in a small-world network.
From the statistical point of view networks may have a characteristic distribution of nodes $P(k)$ as a function of the number $k$ of connections per node. E.g., a small-world network typically has a Gaussian distribution $P(k) \propto \exp((k-k_0)^{2})$. In contrast the scale-free network has a power-law distribution $P(k) \propto k^{\alpha}$. SFNs also occur widely in nature. One example is the network of airports in the US. Recently, Makse et al.~\cite{Song05} discovered that the WWW network and some patterns in biochemistry have something in common with the architecture of snowflakes and trees. Those are self-similar (they look the same when looked upon at different length scales, e.g. by magnifying glasses of different strength). Mathematically this means they are fractals. 
\begin{figure*}[ht]
\vspace{0.0cm}
\begin{center}
\includegraphics[scale=0.6,angle=-90]{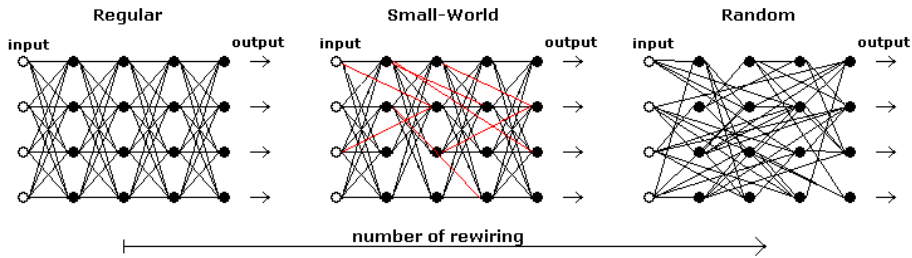}
\vspace{0.0cm}
\end{center}
\caption{Scheme of network connection topologies obtained by randomly 
cutting and rewiring connections, starting from regular net (left) 
and going to random net (right). The SWN is located 
somewhere in between.} 
\label{fig:connectivity}
\end{figure*}
In biological systems, SWNs and SFNs have been found to play a role. For example, food webs of fish in the ocean have been found to be SWNs, another example being protein networks in yeast. In living animals like in the macaque visual cortex the topology of neural connectivity has been found to be SWN~\cite{Sporns00}, likewise in cat cortex~\cite{Sporns00}.
In human brain using functional magnetic resonance imaging, correlations of human brain activity have been measured, representing a network of small-world and scale-free 
architecture~\cite{Chialvo04}. However, it should be pointed out that due to spatial resolution in mm range, these correlations are not representing single neurons, but rather correlations between spots containg many neurons (in the order of $10^{5}$).

This leads to the following important question. If nature seems to employ the SWN and SFN architecture it probably is a sucessful organizational principle in the sense of evolution.  
What advantages may result from such architectures?
First, the network of cortical neurons in the brain has sparse long ranged connectivity, 
which may offer some advantages of small world connectivity~\cite{Laughlin03}.
In studies of small-world neural networks it has turned out that this architecture 
has potentially many advantages. In a Hodgkin-Huxley network the small-world
architecture has been found to give a fast and synchronized response
to stimuli in the brain~\cite{Corbacho00}. In associative memory
models it was observed that the small-world architecture yields the
same memory retrieval performance as randomly connected networks,
using only a fraction of total connection length~\cite{Bohland01}.
Likewise, in the Hopfield associative memory model the small-world
architecture turned out to be optimal in terms of memory storage
abilities~\cite{Labiouse02}. With a integrate-and-fire type of
neuron it has been shown~\cite{Rosin04} that short-cuts permit
self-sustained activity. A model of neurons connected in small-world topology has been used to explain short bursts and seizure-like electrical activity in epilepsy~\cite{Netoff04}.
\begin{figure*}[ht]
\vspace{0.0cm}
\begin{center}
\includegraphics[scale=0.3,angle=-90]{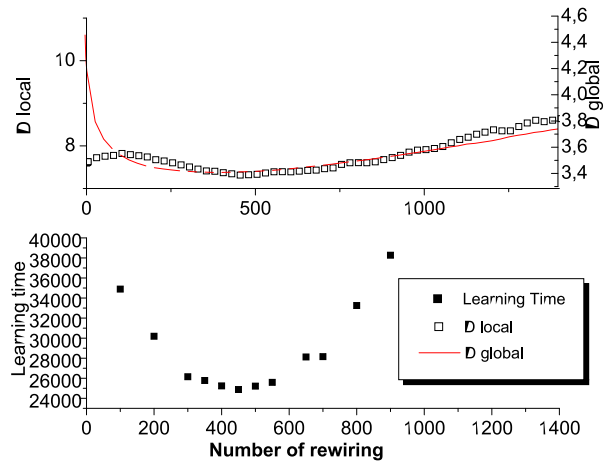}
\vspace{0.0cm}
\end{center}
\caption{Networks with 10 neurons per layer and 10 layers. $D_{local}$, $D_{global}$ and Learning time $T$ versus number $N$ of rewirings. At $N \approx 450$, both $D_{local}$, $D_{global}$ become minimal, i.e. represent SWN connectivity. One observes that the minimum in learning time $T$ also occurs at $N \approx 450$, i.e. coincides with SWN connectivity.}
\label{fig:DLocGlob}
\end{figure*}
The behavior of a small-world network in comparison to a random network and a regular network has been investigated in a model of supervised learning by Simard et al.~\cite{Simard05}. The model is variant of the Perceptron model developped by Rosenblatt~\cite{Rosenblatt62} in the sixties, later being generalized to multi-layer feed-forward networks. The information, representing action potentials
propagating in axons and dendrites of biological neurons, feeds
in forward direction. This can be viewed as a simplistic model of the multi-layer structure of visual cortex. One should note that networks of this type are being used as pattern recognition devices in artificial intelligence.  
E.g., a convolutional network, consisting of seven layers plus one input layer 
has been used for reading bank cheques~\cite{LeCun98}. 
In principle the network is trained with a set of given patterns of information (e.g. representing a sequence of numbers 0 and 1) which should generate a given set of target patterns. Training means that the interneural weights (representing strength of synaptic connections) will be adjusted until the network
generates output patterns as close as possible to target
patterns (hence supervised learning). After training has been achieved the network can be used to analyze new patterns and eventually classify them (in the case of bank cheques or postal letter codes the classes are the alphabetical letters or digital numbers).  
\begin{figure*}[ht]
\vspace{0.0cm}
\begin{center}
\includegraphics[scale=0.6]{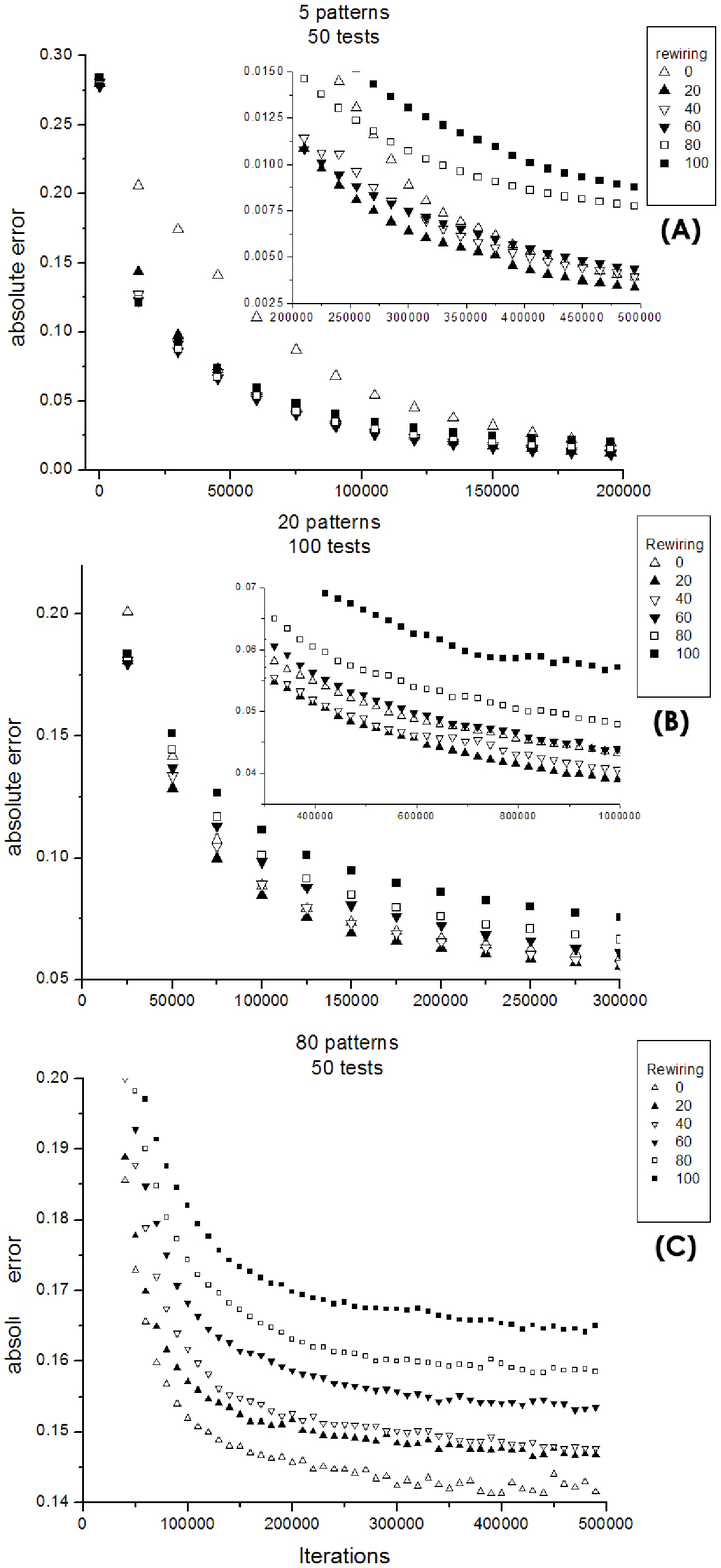}
\vspace{0.0cm}
\end{center}
\caption{Network of 5 neurons per layer and 5 layers. 
Learning of 5 patterns (a), 20 patterns (b) and 80 patterns (c).} 
\label{fig:Function_Pattern}
\end{figure*}
Fig.[\ref{fig:connectivity}] presents a scheme of neural connectivity. One starts from the so-called regular connectivity, where each neuron of a given layer has connections to all neurons of the next subsequent layer in forward direction (left of figure). At the opposite side (right of figure) is shown the so-called random network, where a neuron of a given layer is connected in a random way to neurons of any of the subsequent layers. However, the total number of connections is maintained. Somewhere in between is the SWN, which retains much of the structure of the regular connectivity, but occasionally has some long ranged direct connections to subsequent layers far away. Instead of measuring the network architecture by the functions $C$ and $L$, it is advantageous to use 
the functions of local and global connectivity length $D_{local}$ and $D_{global}$, respectively, introduced by Marchiori et al.~\cite{Latora01,Marchiori00}. 
It has been shown that $D_{local}$ is similar to $1/C$, and $D_{global}$ 
is similar to $L$~\cite{Marchiori00}, although this is not an exact relation. 
Thus, if a network is small-world, both, $D_{local}$ and $D_{global}$ should become
small. Fig.[\ref{fig:DLocGlob}] shows $D_{local}$ and $D_{global}$ as a function of 
the number $N_{rewire}$ of rewiring steps for a network of 10 neurons per layer and 10 layers. One observes that both, $D_{local}$ and $D_{global}$, become small at 
about $N_{rewire} \approx 450$. The learning time is defined as the number of iterations it takes until the training error becomes smaller than a given error tolerance. 
Fig.[\ref{fig:DLocGlob}] shows that the mininum of learning 
time $T$ coincides with minima of $D_{loc}$ and $D_{glob}$, i.e. with SWN connectivity.
\begin{figure*}[ht]
\vspace{0.0cm}
\begin{center}
\includegraphics[scale=0.3,angle=90]{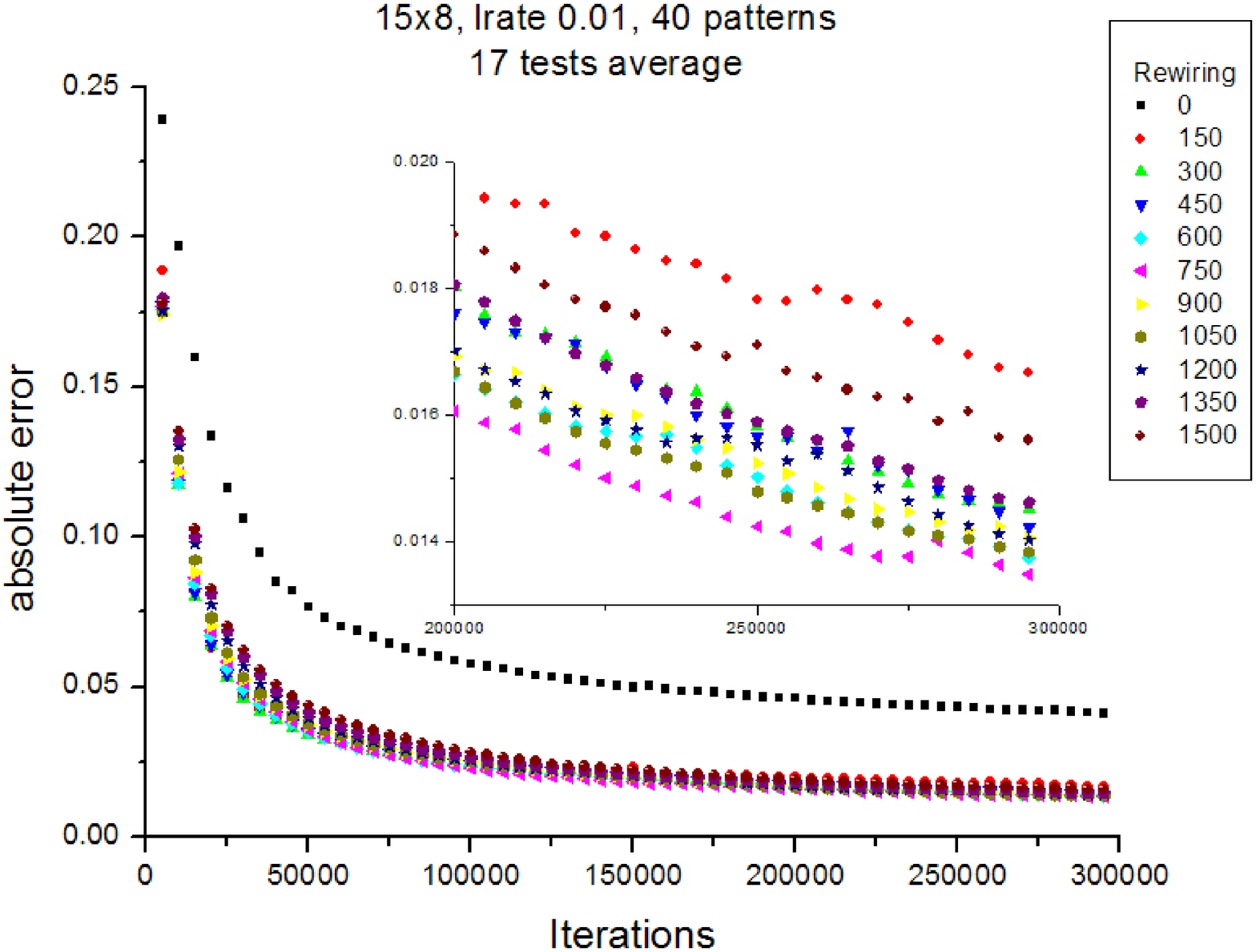}
\vspace{-0.5cm}
\end{center}
\caption{
Network of 15 neurons per layer with 8 layers. 
Network was trained with 40 patterns by 17 statistical tests.
}
\label{fig:Function_Neurons}
\end{figure*}
The effect on learning under variation of the number of layers and the 
effect of rewiring, i.e. introducing short-cuts, has been investigated in a network of 5 neurons per layer with the number of layers varying from 5 to 8. 
Comparison of the regular network (no rewiring) with the network after a few (40) rewirings, which brings it into the regime of small world architecture, the following behavior (not shown here) is found.  
While the regular 5-layer network learns (i.e. the error decreases) with increasing iterations, the regular 8-layer network does not learn at all. After carrying out many iterarations one observes that the network with $N_{rewire}=40$ gives a smaller error compared to the regular network. This holds for 5,6,7 and 8 layers. 
While the improvement over the regular network 
is small for 5 layers, it is major major for 8 layers. 
This indicates that the  learning improvement due to small-world architecture becomes more efficient in the presence of more layers. In the case of 8 layers, 
the error curve for $N_{rewire}=40$ is in the regime where 
$D_{local}$ and $D_{global}$ are both small, 
i.e. close to the small-world architecture.
 
Next the effect on learning when changing the number of learning patterns and the 
rewiring of the network has been investigated for a network of 5 neurons per layer and 5 layers. The results are shown in Fig.[\ref{fig:Function_Pattern}]. 
When learning 5 patterns 
(Fig.[\ref{fig:Function_Pattern}a]) in the domain of few iterations (1000-5000), 
rewiring brings about a substantial reduction in the error compared to the 
regular network ($N_{rewire}=0$) and also to the random network ($N_{rewire}>100$). 
For very many iterations (about 500000) there is little 
improvement with rewirings compared to the regular architecture. For learning 
of 20 patterns (Fig.[\ref{fig:Function_Pattern}b]) the behavior is similar, but the 
reduction of error in the presence of 20 rewirings is more substantial 
for many iterations. When learning 80 patterns Fig.[\ref{fig:Function_Pattern}c] 
the error is large, the system is near or beyond its storage capacity. 
In this case the regular architecture is optimal and rewiring brings no advantage.

The influence of the number of neurons per layer on learning is depicted in 
Fig.[\ref{fig:Function_Neurons}], comparing (a) a network of 5 neurons per layer and 8 layers with a network (b) of 15 neurons per layer and 8 layers. The network has been trained with 40 patterns in both cases. 
In case (a) the SWN case (few rewirings) gives a clear improvement over the regular architecture ($N_{rewire}=0$) 
and the random architecture ($N_{rewire}>100$). 
In case (b) the learning error is shown in Fig.[\ref{fig:Function_Neurons}]. 
One observes that the learning error has a minimum at 
$N_{rewire} \approx 750$, which is close to the minimum of $D_{local}$ and 
$D_{global}$ at $N_{rewire} \approx 830$, i.e. near the SWN regime.  
  
Also generalization has been studied in a network of 5 neurons per layer 
and 8 layers. The network was trained to put patterns into classes. Five classes have been considered, defined as follows. A pattern belongs to class 1, if neuron 1 in the input layer has the highest value of all neurons in this layer. Class 2 corresponds to neuron 2 having the highest value. etc. The classification of 200 patterns achieved by the network has been studied as function of connectivity. It turns out (not shown here) that the network with some rewirings gives improvement over the regular and also random network architecture. In particular, the generalization error has a minimum at  $N_{rewire} \approx 40$, which is in the regime where $D_{local}$ and $D_{global}$ are small, i.e. in the regime of small-world architecture.

In summary of this section, network models have been used to simulate different brain functions like (i) generation of coherent and fast response (in a situation of danger, a fast simultaneous activation of motor neurons is required), (ii) retrieval of corrupted memory patterns, and (iii) supervised learning. In all cases it turned out that SWN connectivity as compared to regular or random connectivity brought about a net advantage. From the purely mathematical point of view, one may argue that a fully connected network (where each neuron is connected to all of the other neurons) may be even better. However, one may easily convince oneself that such neural connectivity is not suitable in brain for reasons of space limitation, but also due to physical limitations like oxygen, blood and energy supply. The human bain actually has very sparse connectivity. Among all varietes of sparse connectivity, the SWN connectivity seems to be a kind of optimum with respect to efficiency in information transfer.

\section{Self-organization of feature maps}
\label{sec:FeatureMaps}
An important principle of neural organization of the brain are the so called feature maps. One example is the organization of the visual cortex. There are many other examples referenced in sect.~\ref{sec:Intro}. The question being addressed here is very simple: How come that the brain recognizes a human face, given the fact that seeing a face means a huge amount of photons coming from all directions into the eye? The answer is not simple at all. It is well known that the retina of the eye has selective neurons, which respond to stimuli of certain types, like lightness, orientation, shape etc. But this is not the same as recognizing a face. The different layers of the visual cortex treat information in the sense that it becomes gradually more complex. In simplified terms, the first layer recognized points and straight lines, in the next layer, it recognizeses balls, cubes, and in further layers a house, a landscape or a face. Recognizing does not mean that a particular image, like a picture of Mont Blanc, would correspond in a one-to-one way to a specific neural configuration. Rather one should imagine that the neural configuration is "close" to a pattern of stimulations. Early models of formation of the visual cortex have been proposed by van der Malsburg~\cite{Malsburg73} and Linsker~\cite{Linsker86}.
The neural organization of the visual cortex, for example in cat kittens, occurs in a short time window (few weeks) after birth. This starts already when their eyes are stil closed.
The idea of the Kohonen model is based on the following physiological observations. First, there is some sort of continuity, e.g., between location of sensor receptor cells and its representation in the sensorimotor cortex,  given by the somatotopic 'homunculus'. It means that receptor cells being close to each other correspond to representations being close to each other in the sensorimotor cortex. Second, different types of response properties may be represented by the same area in the cortex, as has been shown in visual, somatosensory and auditory cortex~\cite{Hubel63,Sur81,Linden03,Friedman04}. Such kind of evidence supports the idea, first suggested by Kohonen~\cite{Kohonen82} that there is a a low-dimensional (2-D) map representing the response of stimuli in the cortex. The target-range map in the bat auditory cortex~\cite{Suga79} is an example of an ordered map of abstract features. As no receptive surface exists for such abstract features, the spatial order of representations must be produced by some self-organizing process in brain.

\begin{figure}
\centering
\begin{tabular}[t]{l l}
\includegraphics[scale=0.45]{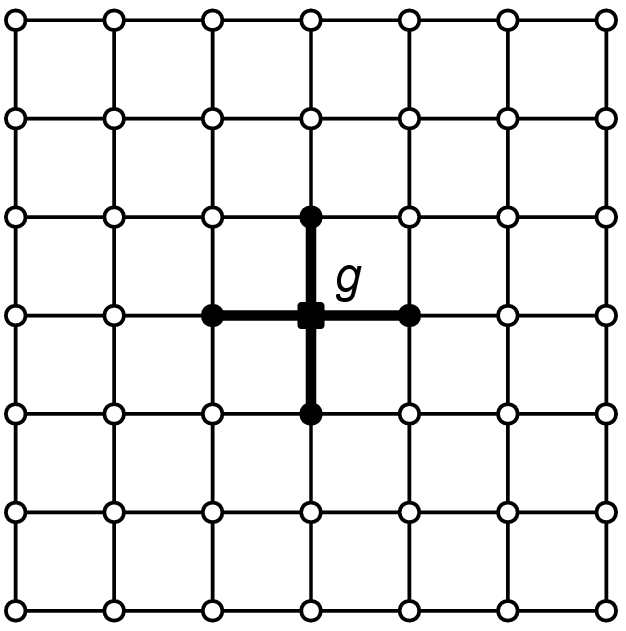} & 
\includegraphics[scale=0.45]{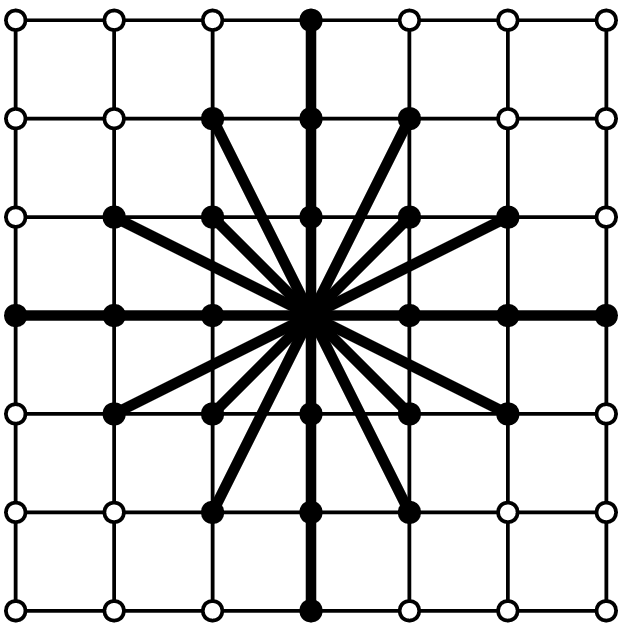} 
\end{tabular}
\caption{Initial regular grid map of neuron topology (neighbors): $V_{g}=1$ (left) and $V_{g}=3$ (right).}
\label{fig:GridMap}
\end{figure}

\begin{figure}
\centering
\begin{tabular}[t]{@{\!\!\!\!\!}c@{\!\!\!\!\!}c} 
\includegraphics[scale=0.30]{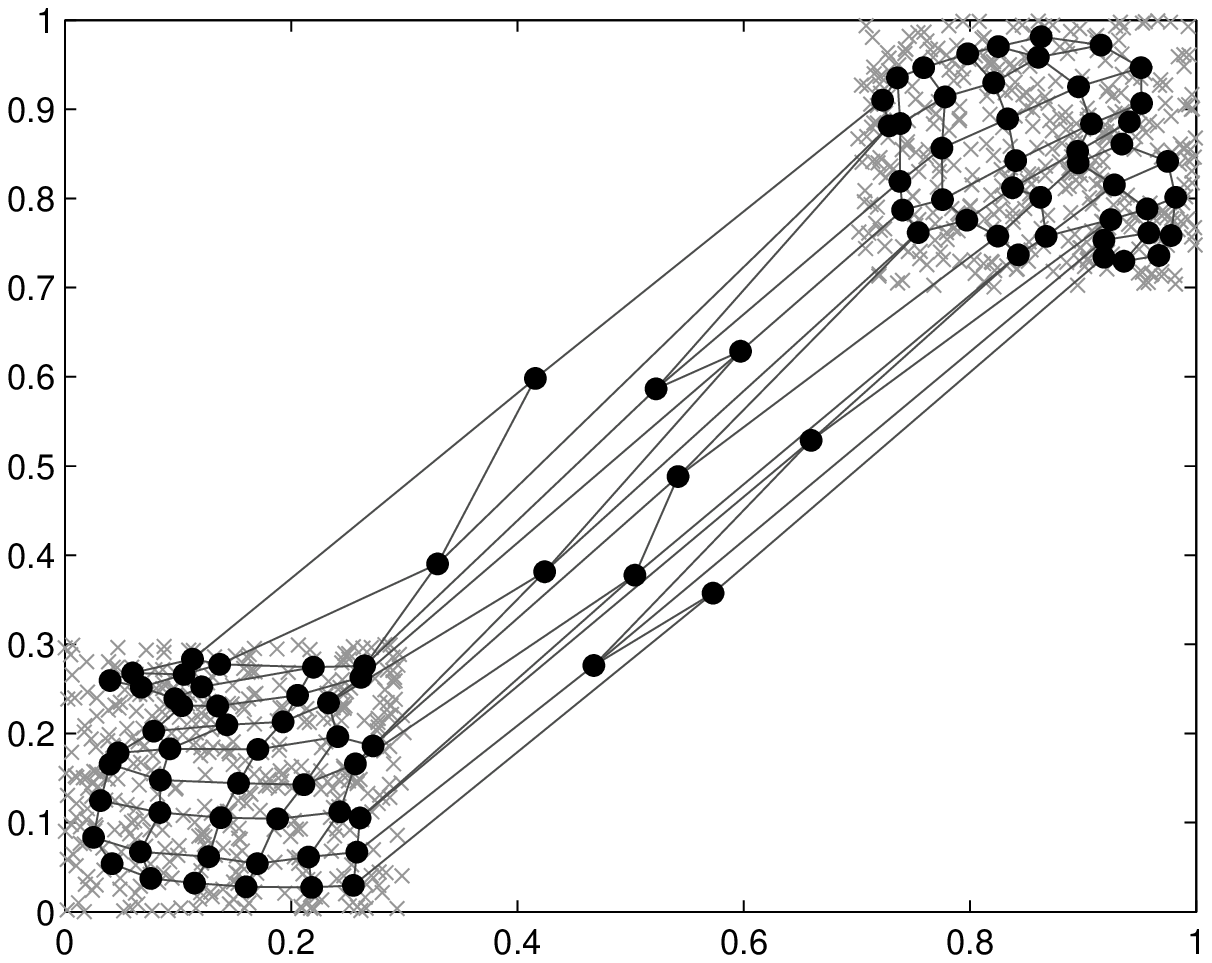} \\
\includegraphics[scale=0.30]{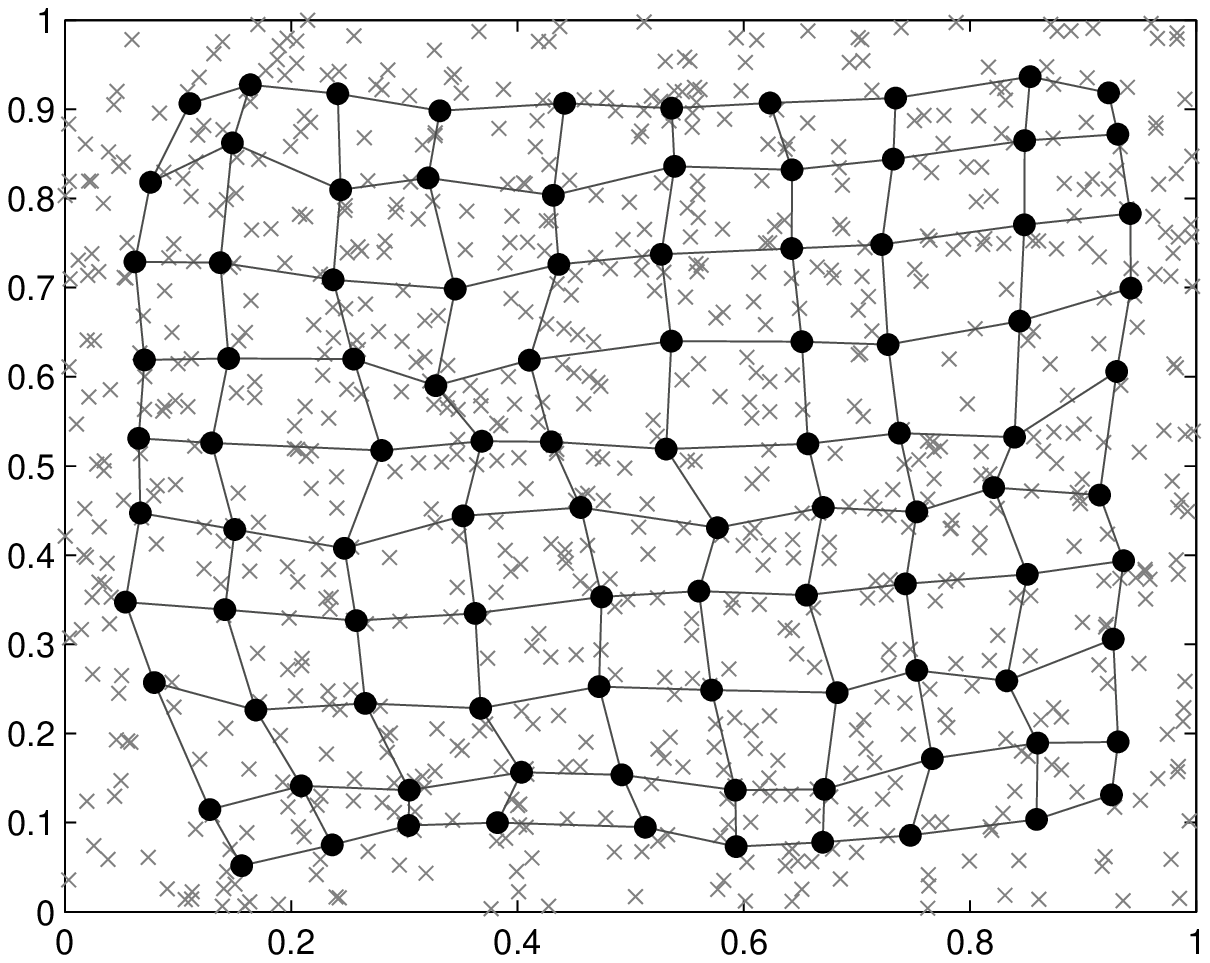} \\
\includegraphics[scale=0.30]{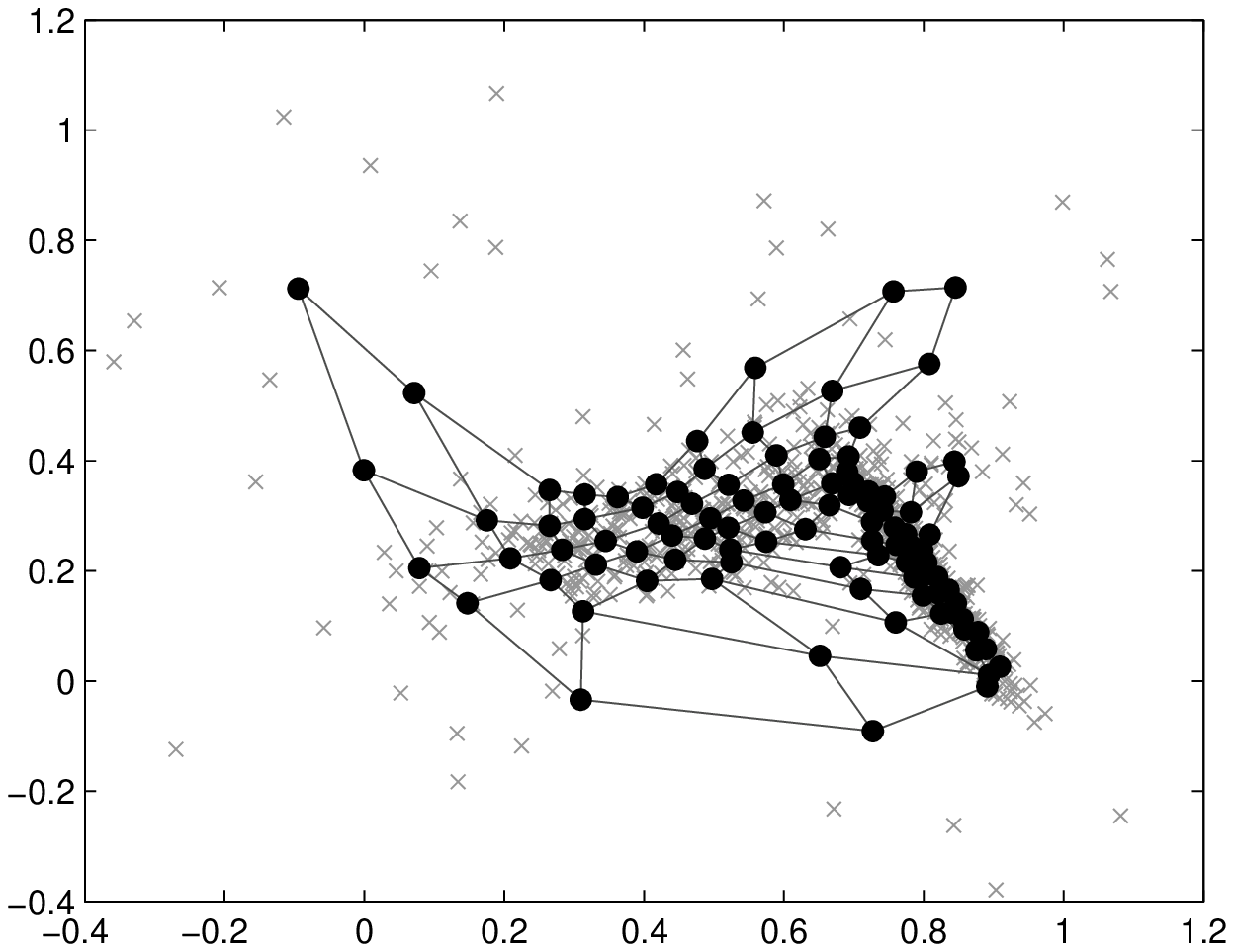} \\ 
\\
\end{tabular}
\caption{Standard Kohonen map, showing stimuli (light crosses) and neurons (bold dots) for 2-D data sets (top=1, middle=2, bottom=3). Lines represent the deformed map at the end of organization.}
\label{fig:FinalStimulNeurMap}
\end{figure}

The standard Kohonen model, or Self-Organized Map (SOM) is a model of formation and self-organization of brain maps. Self-organization means that this process is not supervised by any teacher or control beyond the brain. The model is defined by the following rules.
First there are stimuli. In the example of visual cortex these may be an image given in terms of points having a value of brightness/darkness, a value of an angle (representing the orientation) and a value of color. Thus each point has a triple of qualities, mathematically given by a 3-dimensional vector. In general the stimuli may be viewed as an ensemble of points located at 
$\vec{p}_{q}$, $q=1,\dots,Q$ living in a high-dimensional vector space. Second there is an ensemble of neurons. These neurons are located on a 2-dimensional grid (see Fig. 
\ref{fig:GridMap}).
For each neuron this grid determines its neighbor neurons. The range of the neighborhood 
is parametrized by the parameter $V_{g}$ (Fig. \ref{fig:GridMap}).
In order to establish a connection between stimuli and neurons  
to each neuron has been assigned a weight vector $\vec{w}_i$, $i=1,\dots,N$ living in the same vector space as the stimuli. Using the stimulus vector $\vec{p}_{q}$ and the weight vector of neurons $\vec{w}_i$, one can define a distance between a stimulus and a neuron.
\begin{figure}
\centering
\begin{tabular}[t]{c c}
\includegraphics[scale=0.3]{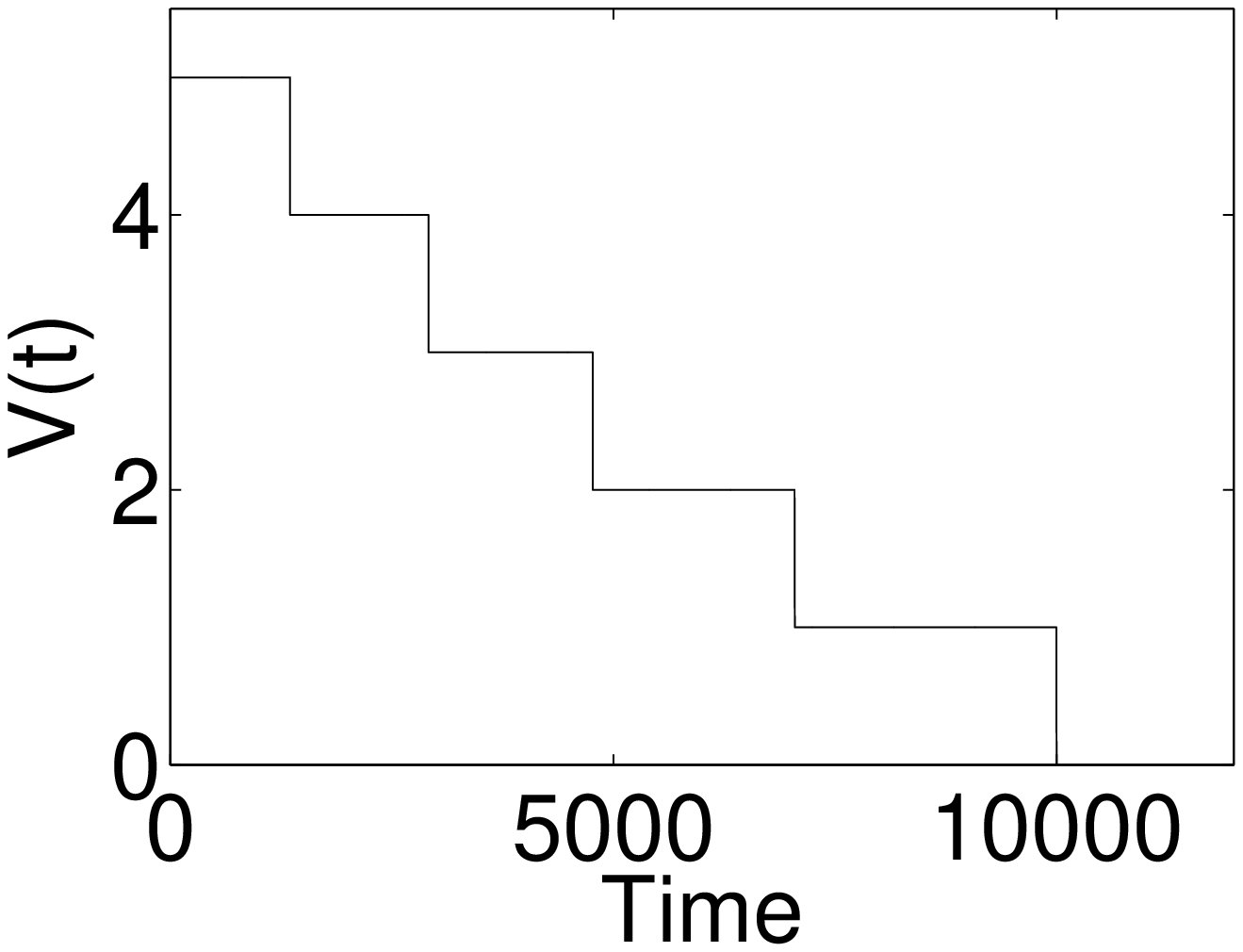} & 
\includegraphics[scale=0.3]{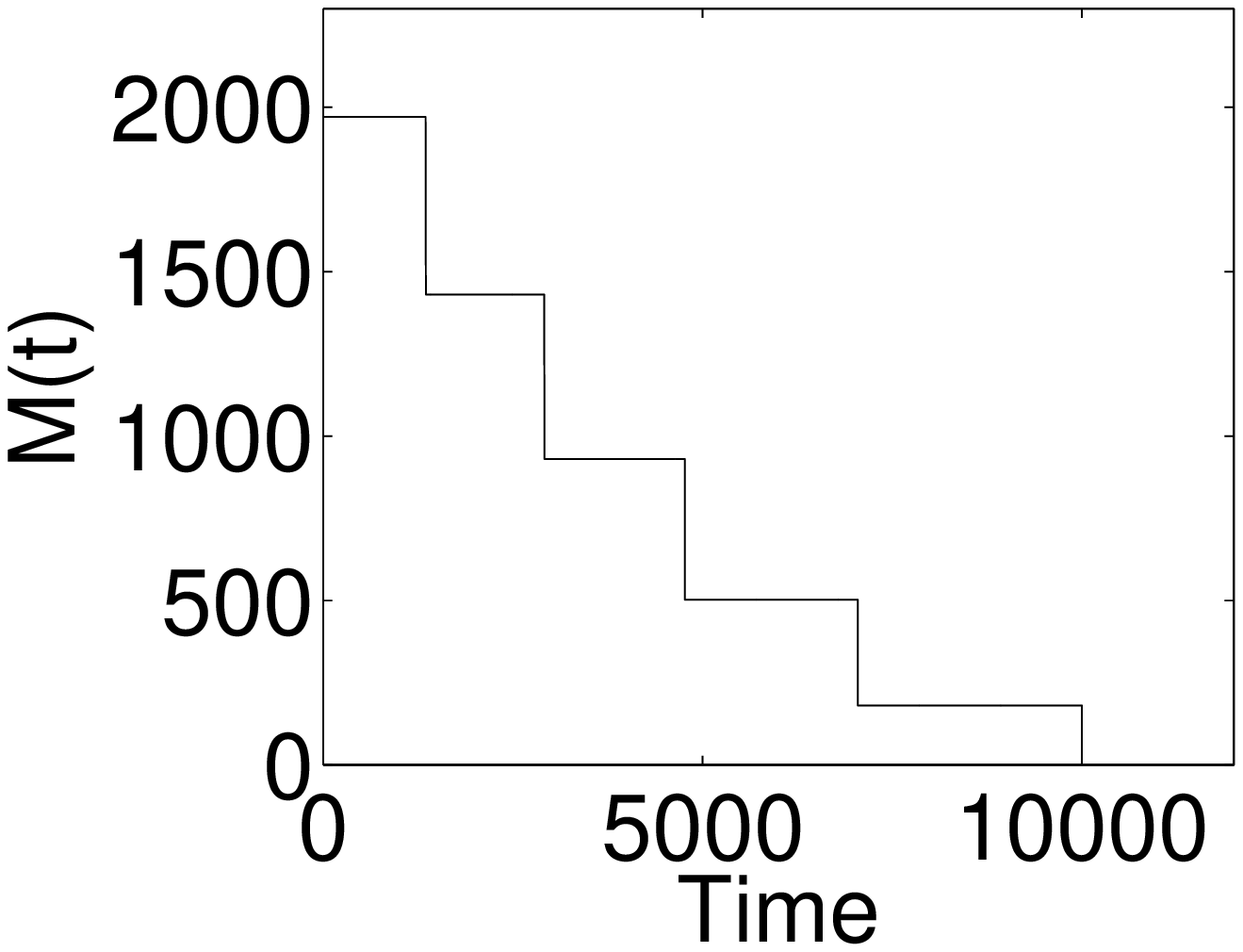}
\end{tabular}
\caption{Standard Kohonen. Relation between neighborhood order $V(t)$ and total number of connexions $M(t)$.}
\label{fig:NeigborhoodConnections}
\end{figure}
The dynamical rules determining the process of organization are the following. At each time $t$, a stimulus $\vec{p}_{q}$ is randomly selected. Then one determines a so-called "winning" neuron, that is the neuron being closest to the stimulus. For such winning neuron, the grid map defines a set of neighbor neurons. Then the weight vectors of the winning neuron and its neighbors are updated according to the rule 
\begin{equation}
\vec{w}_{i}(t+1) = \vec{w}_{i}(t) + 
\eta(t)[\vec{p}_{q}(t)-\vec{w}_{i}(t)], ~ 
\forall i\in V_g (t) ~ .
\label{eq:UpdateWeights}
\end{equation}
The parameter $\eta$ denotes the learning rate. Using this rule, the neuronal map learns the topology of the data set and eventually becomes deformed 
(as shown in Fig. \ref{fig:FinalStimulNeurMap}).
\begin{figure}
\centering
\includegraphics[scale=0.45]{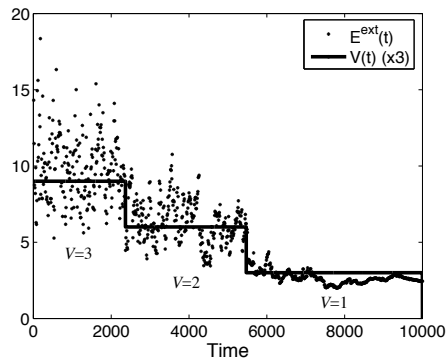}
\caption{Standard Kohonen. Temporal evolution of absolute error $E^{ext}$ and neighborhood order $V$.}
\label{fig:EvolError}
\end{figure}
The order of neighborhood $V(t)$ is initially high, i.e. the map is highly connected. During evolution of organization $V(t)$ decreases gradually 
(Fig. \ref{fig:NeigborhoodConnections}). Also $\eta(t)$  decreases linearly in time to ensure convergence of neuronal weights.
The evolution of the map topology $V(t)$ as well as the learning rate $\eta(t)$ are determined a priori. This conditions the final absolute error of modelization, 
$E^{ext}$, given by 
\begin{equation}
E^{ext}=\sum_{q=1}^{Q} {\min_{i\in \{1,\ldots,N\}} 
\| \vec{p}_{q}-\vec{w}_{i} \|^{2} }  ~ .
\label{eq:erreurabsolue}
\end{equation}
Fig. \ref{fig:EvolError} shows a typical evolution of $E^{ext}$ over time where the neighborhood parameter $V$ decreases in steps from 3 to 1. This means that on average the set of neurons are approaches the set of stimuli.  
Fig.\ref{fig:FinalStimulNeurMap} shows different cases of 2-D data sets of stimuli, the location of neurons at the end of the organization phase and the deformed map of neural connections (which developped from the originally regular neighbor grid map). 

As mentioned previously, the Kohonen model has been shown in a number of cases to be biologically realistic. Obermayer et al.~\cite{Obermayer90,Obermayer92}, Goodhill~\cite{Goodhill93}, Wolf et al.~\cite{Wolf94}, Swindale and Bauer~\cite{Swindale98} have shown that the Kohonen model produces mappings which resemble in detail to real visual cortex maps. In the study of primary visual cortex (V1) of ferret Sur et al.~\cite{Sur05} found a distorsion in the mapping of the visual scene onto the cortex, in agreement with the predictions of the Kohonen model. Aflalo et al.~\cite{Aflalo06} investigated origins of complex organization of motor cortex and found that the Kohonen map contained many features of actual motor cortex in monkey. 

In Sect.~\ref{sec:SWNTopology} it has been shown that Small-World and Scale-Free connectivity seems to play a favorable role in information transfer in neural networks. One may ask if such topology plays a role within the context of the standard Kohonen model. Pallaver~\cite{Pallaver06} has measured the functions $D_{local}$ and $D_{global}$ which, according to Marchiori and Latora~\cite{Marchiori00,Latora01} should both be small in the case of a Small-World network. It turns out that these function are small during most of the time of organization, indicating SWN connectivity, i.e. the network is highly connected at all length scales. Only towards the end of organization, they cease to be small. Then the network looses its SWN character (refinement of spatial scale of learning, formation of separate islands of neurons representing classes of stimuli). The lesson from this is that also in non-supervised self-organization of feature maps, efficient information transfer is very important. Thus SWN connectivity, found to be realized in nature, and being established as a principle of efficient information transfer, turns out to be an important organizational principle also in the Kohonen model of self-organized feature maps.

Finally, let us take a look at neural cell death and pruning of synaptic connections in the brain, which starts very early in a post-natal 
period~\cite{Shepherd94,Kandel95}. This phenomenon has been looked at from the point of view of the Kohonen model. Pallaver~\cite{Pallaver06} has compared the error in organization, once by considering only pruning of synaptic connections (which in the Kohonen model is built in via reducing the neighborhood size $V$), and contrasted it to the case, where reconnections where allowed. The latter option led to a larger error and deterioration of organization. The pruning of connections appears to be an advantage because the progressive independence of neurons leads to a better precision in the local placement of neural weights. This gives a possible explanation for biologically observed pruning of synaptic connections during organization in the brain of mammals in a short postnatal period, which is another important organizational principle.

\section{Outlook}
In this article we have discussed organizational principles in the brain looked upon from the perspective of information transfer. We considered the case of $1/f$ frequency scaling observed in EEG's in humans and LFP's in cat and discussed possible 
underlying mechanism. In particular, the model of Self-Organized Criticality has been considered as a candidate model for such mechanism. There is evidence in favor as well as evidence to the contrary to this. Second we discussed self-organization in the context of feature maps. Here the Kohonen model or variants of have been applied and compared to biological observations. In many instances the Kohonen model was found to be biologically relevant and in some cases to give predictions in quite precise agreement with experiments. Finally, we discussed neural network models of the brain with respect to connectivity architecture of Small-World and Scale-Free type. 
There is some evidence pointing to the possibility that such architecture is realized in cat, macaque and humans. Model studies have shown that such architectures optimize cognitive tasks, like memory restoration, or learning, but also help to coordinate collective responses, like, e.g. fast coherent activations of motor neurons in case of an emergency. 

The matter discussed in the above is mainly based on the dynamics of neurons. Presently, there is much focus on the role of glial cells, which are known to form its own network, talking to each other via chemical messangers, but also to interact with neurons. Thus it may turn out that glial cells are very important for organization and information transfer in the brain. For example, a neural network model involving glial cells, in contrast to networks of neurons only, was found to be more stable.        
Many exciting and open questions lie ahead.
\\

%
%
H.K. has been supported by NSERC Canada.

\end{document}